\documentclass[mathleft
]{an}
\usepackage{times}
\usepackage{txfonts}
\usepackage{natbib}
\usepackage{longtable}
\usepackage[pdftex]{graphicx}

\begin{document}

% The following seven commands are intended for editorial usage and should be ignored by
% the author(s).
\Pagespan{789}{}% Document's page range. 
% If second parameter is left empty, the last page is computed automatically.
\Yearpublication{2006}%
\Yearsubmission{2005}%
\Month{11}%   
\Volume{999}%  
\Issue{88}% 
% \DOI{This.is/not.aDOI}% 

\title{AGN jets under the microscope: A divide?}
%,\thanks{Data from STELLA}}

\author{M. Karouzos\inst{1,2}\fnmsep\thanks{Corresponding author:
  \email{mkarouzos@astro.snu.ac.kr}\newline}
%Example 
%for footnote, note the usage of the \texttt{fnmsep}
%command as separator between institute number and footnote mark} 
\and  S. Britzen\inst{2}
\and A. Witzel\inst{2}
\and A.J. Zensus\inst{2,3}
\and A. Eckart\inst{3,2}
}
\titlerunning{AGN jets under the microscope}
\authorrunning{M. Karouzos et al.}
\institute{
CEOU/Department of Physics \& Astronomy, Seoul National University, Gwanak-gu, Seoul 151-742, Korea
\and
Max-Planck-Institut f\"ur Radioastronomie, Auf dem H\"ugel 69, 
D-53121 Bon, Germany
\and 
I. Physikalisches Institut, Universit\"at zu K\"oln, Z\"ulpicher Str. 77, 50937 K\"oln, Germany
}

\received{}
\accepted{}
\publonline{later}


\abstract{A new paradigm for active galactic jet kinematics has emerged through detailed investigations of BL Lac objects using very long baseline radio interferometry. In this new scheme, most, if not all, jet components appear to remain stationary with respect to the core but show significant non-radial motions. This paper presents results from our kinematic investigation of the jets of a statistically complete sample of radio-loud flat-spectrum active galaxies, focusing on the comparison between the jet kinematic properties of BL Lacs and flat-spectrum radio-quasars. It is shown	 that there is a statistically significant difference between the kinematics of the two AGN classes, with BL Lacs showing more bent jets, that are wider and show slower movement along the jet axis, compared to flat-spectrum radio-quasars. This is interpreted as evidence for helically structured jets.}

\maketitle

\section{Introduction}
Black holes have been at the center of astronomical research for many years, being one of the most prominent predictions of the theory of General Relativity. There is a large amount of indirect evidence supporting that most, if not all, galaxies host a supermassive black hole (SMBH) at their centers (e.g., \citealt{Blandford1986}).  It is widely believed that this SMBH is the main constituent of activity in active galactic nuclei (AGN) (for a review see e.g., \citealt{Begelman1984}). Given their very luminous nature, AGN are therefore some of the best probes that we have today to study the properties and evolution of SMBH out to the earliest times of the Universe. It is believed that AGN are made up of a number of building blocks including a central SMBH, an accretion disk, an obscuring molecular screen (usually referred to as a torus), ionization regions producing broad and narrow emission lines, and in some cases powerful collimated outflows, known as AGN jets (e.g., \citealt{Antonucci1993}, \citealt{Urry1995}).

\begin{figure}
\includegraphics[width=0.49\textwidth]{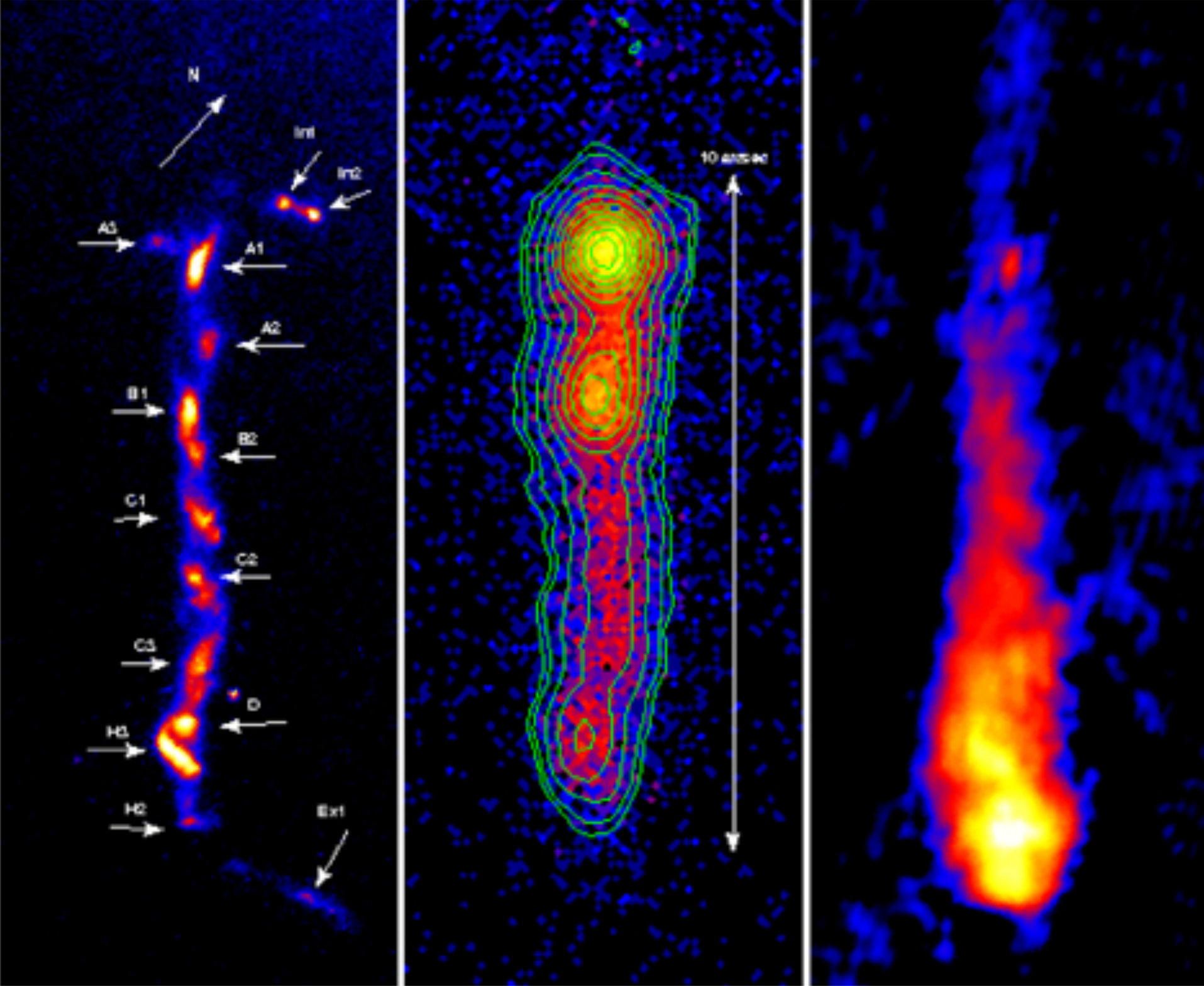}
\caption{The active galactic nucleus 3C273 and its jet observed in three different wavelength regimes, the optical (left panel, \textit{HST}), the X-ray (middle panel, \textit{Chandra}), and the radio (right panel, MERLIN). Credit: NASA/STScI, NASA/CXC, MERLIN.}
\label{fig:AGNmultijet}
\end{figure}

Jets are found in a plethora of astrophysical environments, among which, young stellar objects, massive X-ray binaries, possibly pulsars, $\gamma$-ray bursts, and AGN. They\linebreak have been observed for the first time in the optical in the nearest active galaxy to us, M87 (\citealt{Curtis1918}).  Given their ubiquity, astrophysical jets research has been a booming field of astronomy, especially after the advent of high-\linebreak resolution imaging instruments and techniques (e.g., very long baseline interferometry, VLBI). Although observed in the minority of active galaxies ($\sim15\%$; \citealt{Krolik1999}), extragalactic jets are some of the most pronounced morphological features in AGN research. Their, presumably direct, connection to the active core and the SMBH residing there, makes them invaluable tools in the effort to characterize the properties and the underlying physics of activity in galaxies. Although observable at different wavelengths (see Fig. \ref{fig:AGNmultijet}), VLBI observations enabled the direct imaging of AGN jets at the highest possible resolutions and thus the detailed study of their properties. 

VLBI techniques have allowed the mapping of jets in the radio. A result of this was the discovery of the knotty nature of AGN jets, observed as a consecutive series of brightness maxima and minima. Observations at different times revealed that these brightness maxima actually move. The investigation of extragalactic jet kinematics had begun. One of the most prominent discoveries, related to jet kinematics, was that of the superluminal movement of these jet components, which was first theoretically predicted by\linebreak \citet{Rees1966}. The effect is a combination of relativistic expansion speeds (close to the speed of light) and the projected geometry onto the plane of the sky. Superluminal movement was indeed detected, first indirectly (\citealt{Whitney1971}), and then by direct imaging (\citealt{Pearson1981}). Superluminally moving components have become the staple of blazars, i.e., radio-loud AGN seen at the smallest viewing angles to their jet axis.

\begin{figure}
\begin{center}
\includegraphics[width=0.45\textwidth]{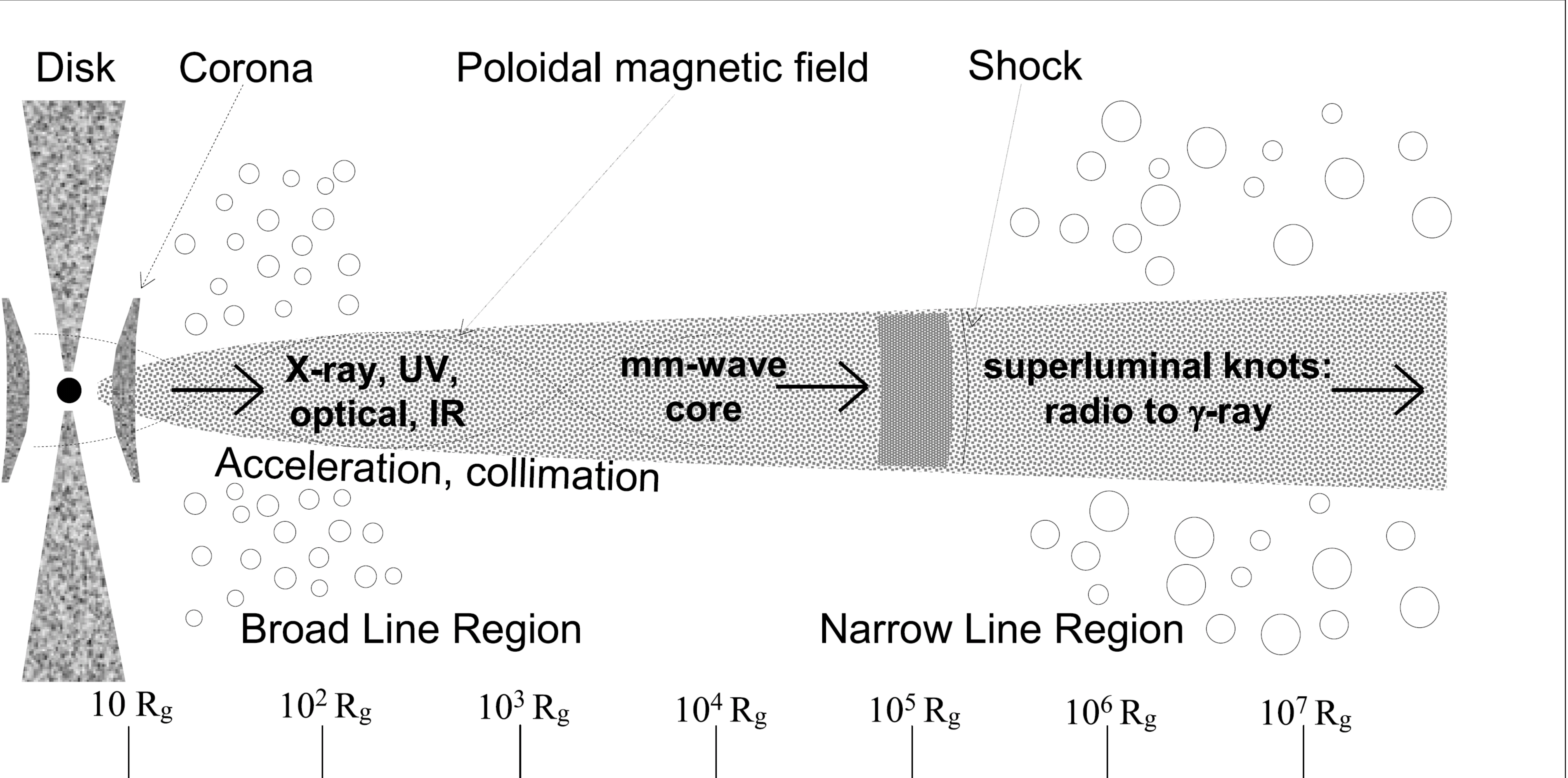}
\caption{An illustration of the AGN core, comprising the SMBH, the accretion disk, line emission regions, and the jet. Different jet regions and the effects observed in them are presented. Illustration reproduced from \citet{Lobanov2007} (adapted from \citealt{Marscher2005}).}
\label{marscher}
\end{center}
\end{figure}

Jet kinematics, as investigated through the study of distinct components, is currently explained in terms of the\linebreak shock-in-jet model (e.g., \citealt{Marscher1985}, also see Fig. \ref{marscher}), where the observed jet knots are manifestations of shocks propagating at relativistic speeds down the jet. As discussed above, beaming and projection effects regulate the observed properties of the jets. Analysis of samples of active galaxies containing large number of sources, selected following stringent criteria, has been of fundamental importance to this end (e.g., \citealt{Ghisellini1993}; \citealt{Vermeulen1994}; \citealt{Vermeulen1995}; \citealt{Taylor1996}; \citealt{Hough2002}; \citealt{Lister2005}; \citealt{Britzen2007a}). There has been a continuous effort to distinguish whether the different types of radio-loud active galaxies (e.g., quasars, BL Lacs, etc.) and their jet properties are an effect of different viewing angles, or whether these objects have intrinsically different properties. The current paradigm is that the differences observed can be attributed to geometrical effects, although some indications to the contrary also exist (e.g., \citealt{Gabuzda1995}; \citealt{Gabuzda2000}; \citealt{Britzen2009}; \citealt{Britzen2010b}). For example, using a sample of 39 superluminal sources,\citet{Ghisellini1993b} find no appreciable difference between the Doppler factors distributions of BL Lacs and quasars, with radio galaxies showing smaller values.

\section{A different paradigm?}
\label{sec:1803}
\begin{figure*}
\begin{center}
\includegraphics[width=0.7\textwidth]{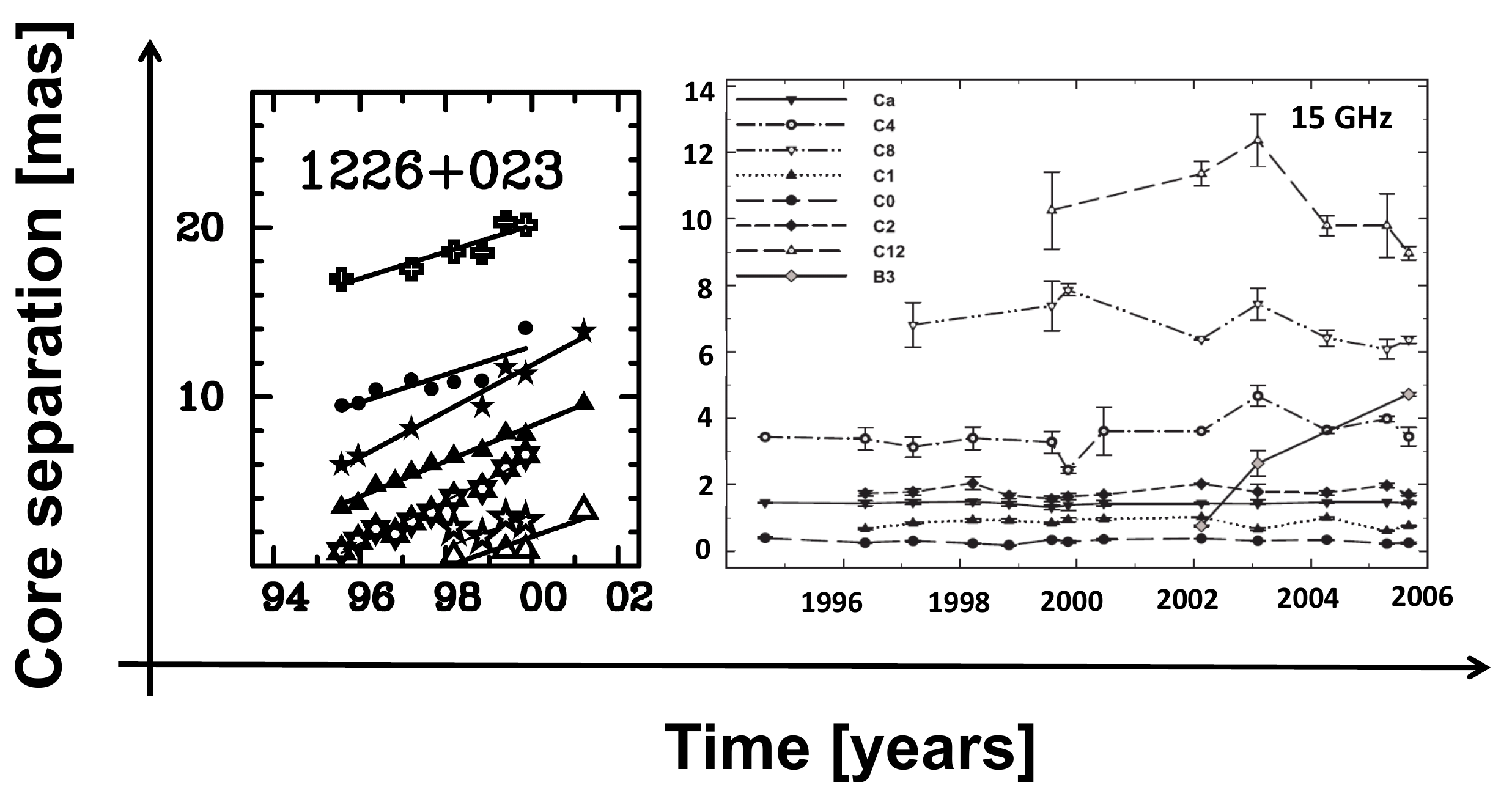}
\caption{A comparison of the kinematic properties of two flat-spectrum radio-loud AGN, 1226+023 (left) and S5 1803+784 (right). The plots show the core separation of the sources' jet components as a function of time. 1226+023 shows clear outward motion at superluminal speeds, while S5 1803+784 shows most components remaining at almost constant core-separations. 1226+023 plot reproduced from \citealt{Kellermann2004} and S5 1803+784 from \citealt{Britzen2010}. For both plots, VLBA data at 15 GHz are used.}
\label{comparison}
\end{center}
\end{figure*}

Previous results concerning both individual sources as well as statistical samples have shown that flat-spectrum AGN predominantly show apparent superluminal motions, usually interpreted in terms of radial, ballistic motions. Contrary to this, recent detailed studies of a number of BL Lac objects suggest a set of somewhat different kinematic properties (see Fig. \ref{comparison}).

Through the (re-)analysis of more than 90 epochs of global VLBI and VLBA data of the BL Lac source S5\linebreak 1803+784, a new kinematic scheme for that source was revealed (\citealt{Britzen2010}). All components in the inner part of the jet (up to 12 mas) appear to be quasi-stationary in terms of their core separation. This behavior is seen at all frequencies studied by the authors (1.6 - 15 GHz). In contrast to this, the components show strong changes in their position angles, implying a dominant component of motion perpendicular to the jet axis.

\citet{Britzen2010} also studied the jet ridge line properties of S5 1803+784. A jet ridge line at a given epoch is defined as the line that linearly connects all component positions at that epoch (for some examples of jet ridge lines, see Fig. \ref{fig:ridge_lines_example}). The jet ridge line of S5 1803+784 is found to change in an almost periodic manner, starting by resembling a straight line, evolving into a sinusoid-like pattern, and finally returning to its original linear pattern, although slightly displaced from its original position. The authors calculate a period of $\sim 8.5$ years for the evolution of the jet ridge line.

Finally, the authors find that the jet changes its apparent width (in a range between a few and a few tens of degrees) in an almost periodic way, with a timescale similar to the one found from the evolution of the jet ridge line. All of the above properties support a new kinematic scheme for 1803+784, where components follow oscillatory-like trajectories, with their movement predominantly happening\linebreak perpendicular to the jet axis rather than along it. In addition, the jet appears at times to form a wide channel of flow, while changing its width considerably across time.

However, S5 1803+784 is not the only exception. The sources S5 0716+714 (\citealt{Britzen2009}) and PKS\linebreak 0735+178  (\citealt{Britzen2010b}), both classified as BL Lac objects, show similar behavior, with strong non-radial motions discovered in their jets. Although not as pronounced as in S5 1803+784, but also lacking the unprecedented number of VLBI epochs of S5 1803+784, both sources show jet components in their inner-jets that appear quasi-stationary with respect to the core. In addition, PKS 0735+178 appears to interchange between the classical blazar-like behavior of superluminally moving components and the behavior seen in S5 1803+784. PKS 0735+178 might be the key to our understanding of these non-radial motion dominated jets and potentially implies a geometrical effect as the origin of the observed properties.

\subsection{Motivation}
Under the unification scheme of active galaxies (e.g., \citealt{Antonucci1993}; \citealt{Urry1995}), both BL Lac objects and flat-spectrum radio-quasars (FSRQs) are believed to be active galaxies for which the viewing angle to their jet axis is very small, leading to strong relativistic effects. In contrast to the latter, BL Lac objects are additionally characterized by the absence of strong emission lines. Assuming similar viewing angle distributions, in the last several years the common classification of blazar has often been used to describe members of either class, also in terms of their jet properties. However, this phenomenological unification of FSRQs and BL Lacs comes into question in light of the recent investigation of sources like S5 1803+784, PKS 0735+178 and S5 0716+714. The question arises\linebreak whether the peculiar kinematic behavior seen in these objects is revealed due to the unprecedented richness of the datasets available, and is therefore relevant for all flat-\linebreak spectrum radio-AGN, or whether BL Lacs are characterized by a genuinely different set of kinematic properties.

We investigate the relevance of the above described,\linebreak new, kinematic scheme for flat-spectrum radio-AGN, while investigating the apparent divide between BL Lac and\linebreak FSRQ jet kinematics. We use the Caltech Jodrell-bank flat-spectrum (CJF) sample to statistically investigate and assess the similarity, or divergence, of the kinematic and morphological properties between the two distinct sub-samples of FSRQs and BL Lac objects in the CJF. We want to test whether jet components of BL Lac objects do show slower apparent speeds with respect to their cores compared to FSRQs. Furthermore we are interested in the phenomenon exhibited in S5 1803+784 of a, at times, very wide jet, with a jet ridge line intermittently resembling a sinusoid-like morphology. To do that we use tools that extract information from the jet ridge line of the sources and characterize therefore the jet globally. This allows for a investigation that is mostly independent of component modeling and component cross-identification. 

The paper is organized as follows: in Sect. \ref{sec:cjf} we describe the CJF sample, in Sect. \ref{sec:data} we shortly describe the data and tools used, in Sect. \ref{sec:analysis} we present the analysis of our data and the results, and in Sect. \ref{sec:discussion} we discuss our results and give some conclusions. Throughout the paper, we assume the cosmological parameters $H_{0}=71$ $\mathrm{km s}^{-1} \mathrm{Mpc} ^{-1}$, $\Omega_{M}=0.27$, and $\Omega_{\Lambda}=0.73$ (from the first-year WMAP observations; \citealt{Spergel2003}).

\section{The CJF sample}
\label{sec:cjf}

\begin{figure}
\begin{center}
  \includegraphics[width=0.3\textwidth,angle=0]{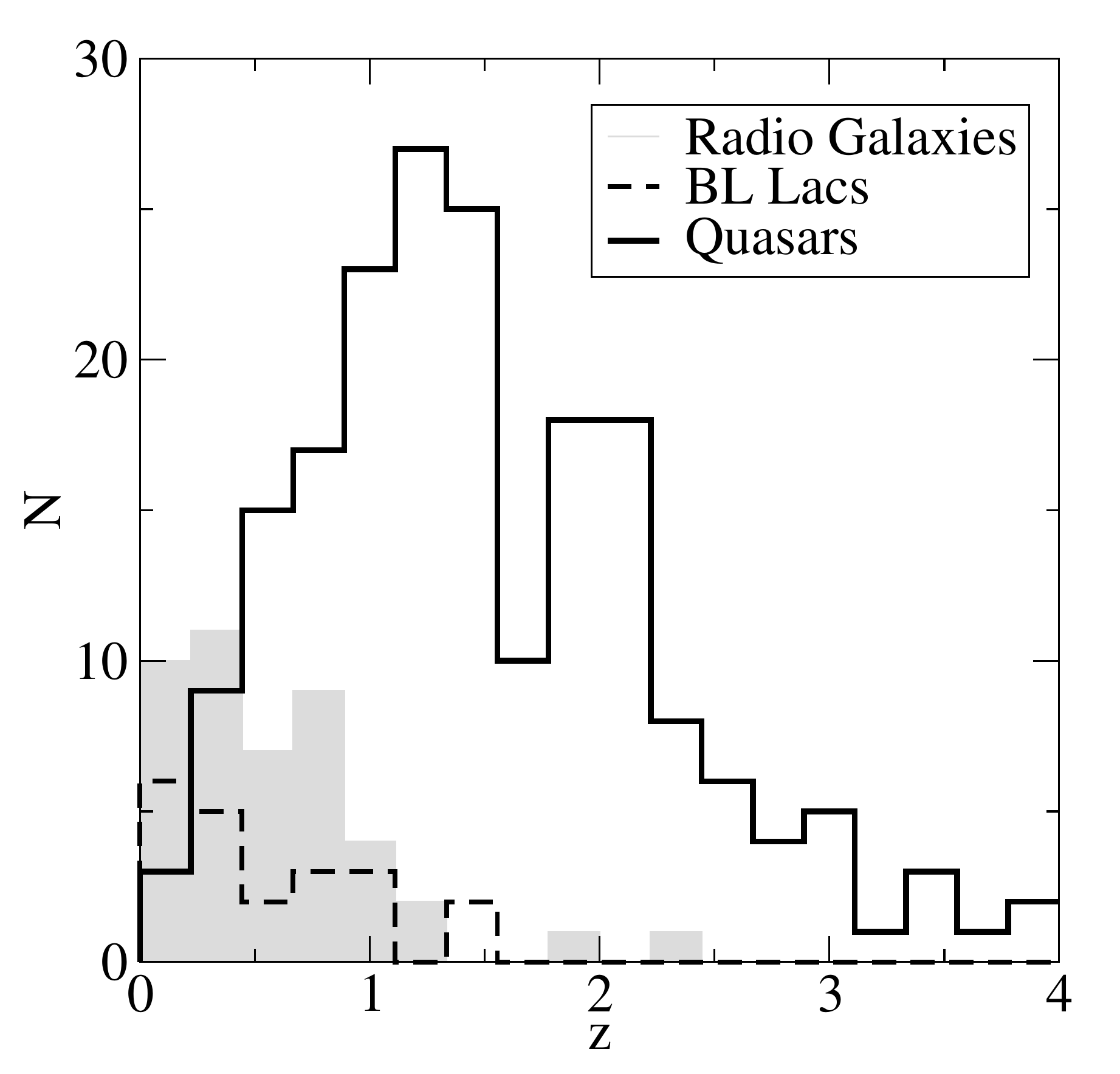}
  \caption{Redshift distribution of radio galaxies (grey blocks), BL Lacs (dashed line), and FSRQs (solid line) in the CJF sample.}
  \label{fig:redshift_histo}
\end{center}
\end{figure}

The CJF sample (\citealt{Taylor1996}) is a flux-limited radio-selected sample containing 293 radio-loud active galaxies selected from three different samples (see Table \ref{tab:cjfproperties} and\linebreak \citealt{Britzen2007a}). The sources span a large redshift range (see Fig. \ref{fig:redshift_histo}), the farthest object being at a redshift $z=3.889$ (1745+624; \citealt{Hook1995}) and the closest at $z=0.0108$ (1146+596; \citealt{deVaucouleurs1991}). The average redshift of the sample is $z_{avg}=1.254$, $z_{BL Lac,avg}=0.546$, $z_{RG,avg}=0.554$, and $z_{FSRQ,avg}=1.489$ for BL Lacs, radio galaxies, and FSRQs, respectively. All the objects have been observed  with the VLBA and/or the global VLBI network. Each source has at least 3 epochs of observations (with a maximum of 5 epochs) and has been imaged and studied kinematically (\citealt{Britzen1999}; \citealt{Britzen2007a}; \linebreak\citealt{Britzen2008}). The X-ray and gamma-ray properties have been studied and correlated with their VLBI properties (\citealt{Britzen2007b} and \citealt{Karouzos2011a}, respectively). The evolution of active galaxies, in the context of the merger-driven evolution scheme (e.g., \citealt{Hopkins2006}), has also been investigated with the help of the CJF, identifying candidate CJF sources in different evolutionary stages, including new binary black hole candidates (\citealt{Karouzos2010}).

\begin{table}[htp]
\caption{CJF sample and its properties.}
\label{tab:cjfproperties}
\begin{tabular}{l c}
\hline\hline
\textbf{Frequency (MHz)}      &  4850 \\
\textbf{Flux lower limit @5 GHz}    &  350mJy\\
\textbf{Spectral Index}      &  $\alpha_{1400}^{4850}\geq -0.5$ \\
\textbf{Declination}         &  $\delta\geq 35^{\circ}$  \\
\textbf{Galactic latitude}   &  $|b|\geq 10^{\circ}$ \\
\textbf{\# Quasars}           &  198 \\
\textbf{\# BL Lac }           &  32 \\
\textbf{\# Radio Galaxies}    &  52 \\
\textbf{\# Unclassified}      &  11  \\
\textbf{\# Total}             &  293 \\
\hline
\end{tabular}
\end{table}

\section{Data}
\label{sec:data}

The work presented here is heavily based on the kinematic analysis of the CJF sample (\citealt{Britzen2007a}; \citealt{Britzen2008}). Five CJF sources were initially excluded from any further analysis due to problematic observations (0256+\linebreak424, 0344+405, 0424+670, 0945+664, 1545+497) and\linebreak therefore are not considered here. In total, 288 sources are considered and analyzed in the following sections. Of these, according to the optical classification from \citet{Britzen2007a}, 196 are classified as quasars, 49 as radio galaxies, 33 as BL Lac objects , and 10 are not classified.

Due to the scope of the CJF program, the identification and analysis of pc-scale jet component kinematics has focused on the part of the jet that is beamed towards us. For a number of sources, several components belonging to the counter-jet have been identified. However, for these sources cross-identification of the counter-jet components over\linebreak epochs has not been carried out. For this reason, and given the nature of the analysis that we undertook (see below), we have excluded all counter-jet components in the following investigation.

The tools that we use for the analysis of the CJF jet ridge lines, in the context described in Sect. \ref{sec:1803}, are described below:
\begin{itemize}
\item Monotonicity Index, M.I.
\item Apparent Jet Width, dP
\item Apparent Jet Linear Evolution, $\Delta\ell$
\end{itemize}
We note that all the above measures, as their names imply, refer to values projected onto the plane of the sky. For the following sections, we adopt the basic assumption that BL Lacs and FSRQs are seen at the smallest viewing angles. As we are interested in the comparison between FSRQs and BL Lacs, deprojection of the jet properties investigated here is not critical. For the following analysis, and for the sake of brevity, we shall drop the characterization of ``apparent'' for each of these values, although this will be implied throughout.

\begin{figure*}[ht]
\begin{center}
  \includegraphics[width=0.5\textwidth,angle=0]{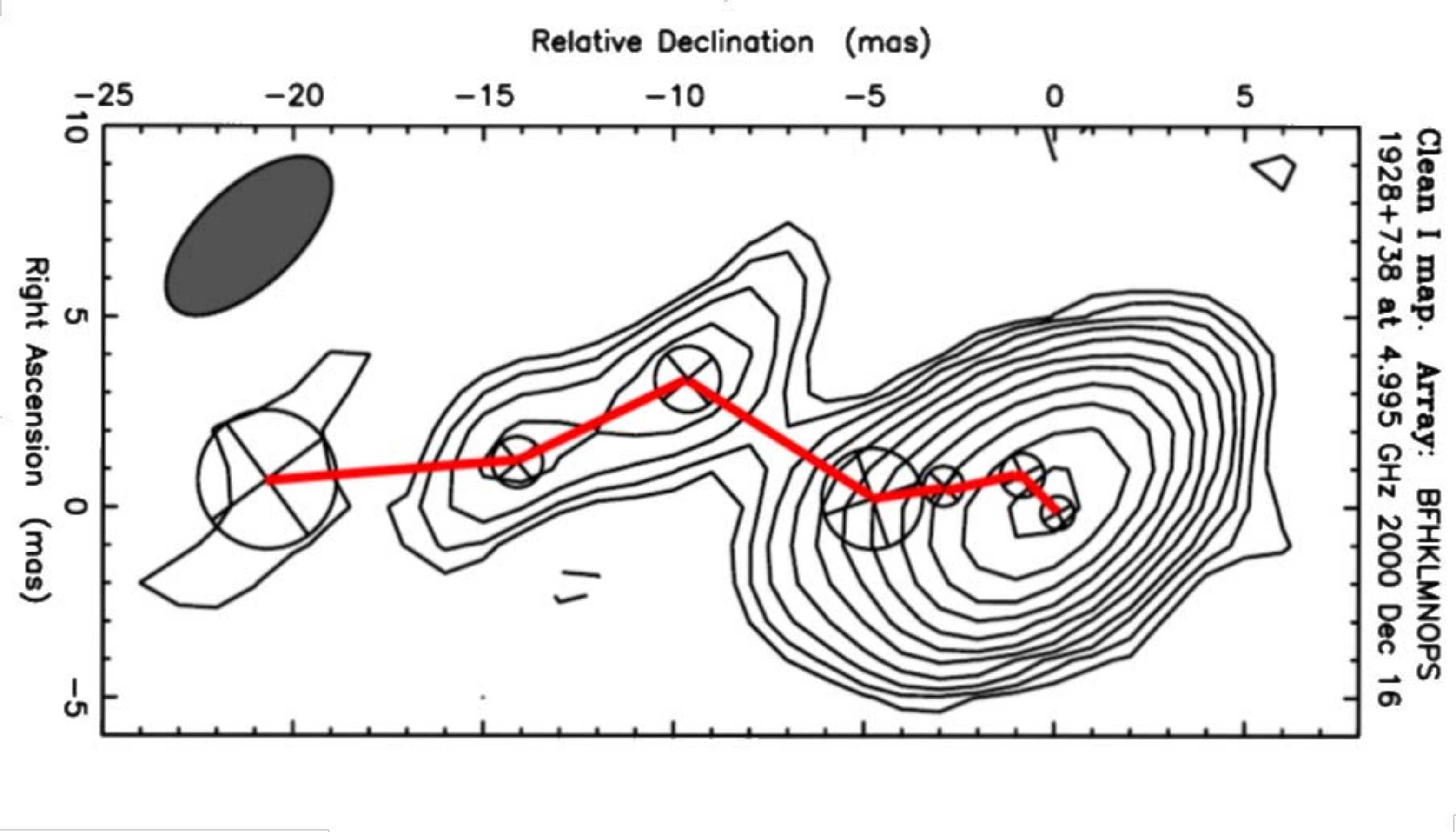}\hspace{20pt}
  \includegraphics[width=0.4\textwidth,angle=0]{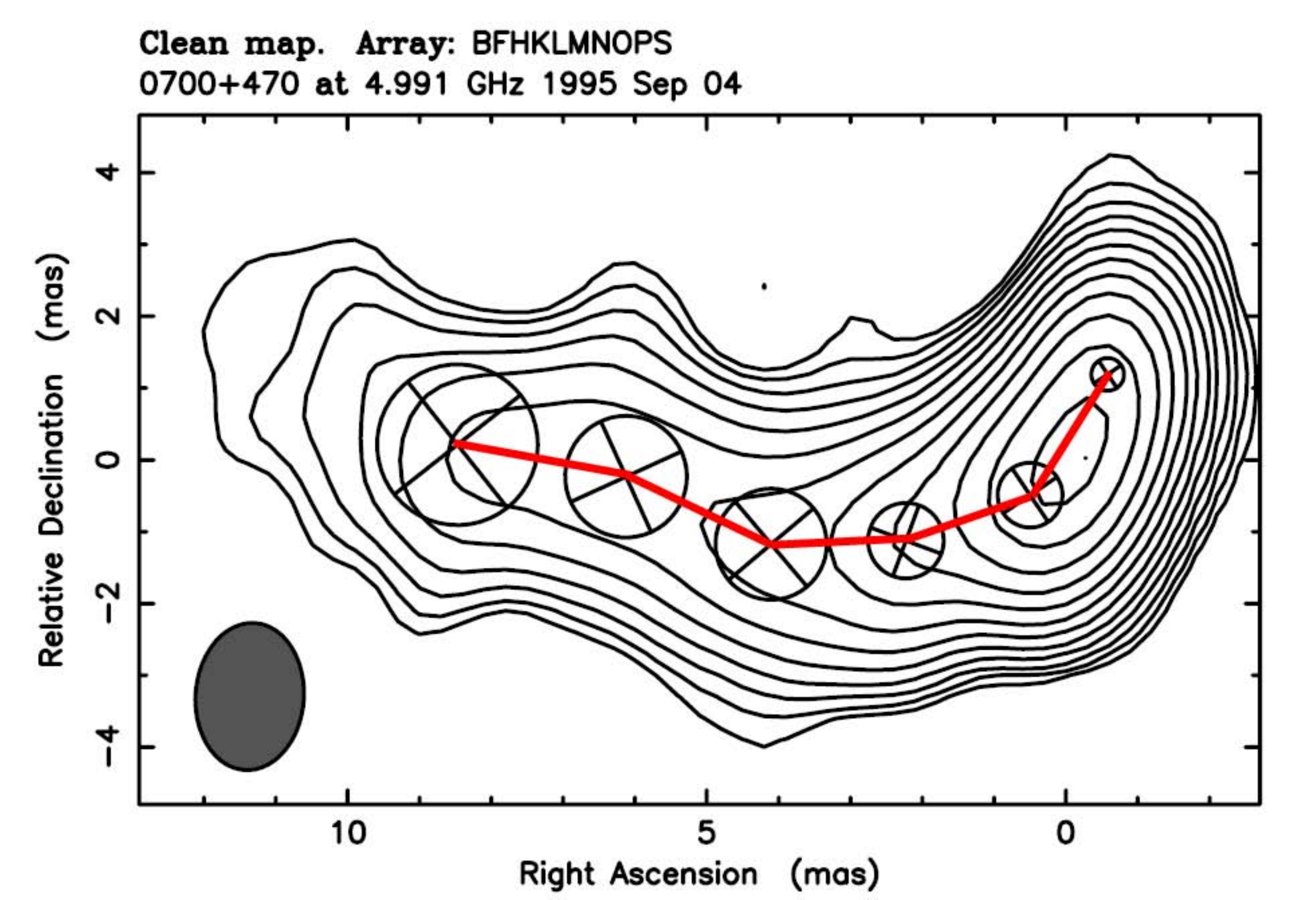}
  \caption[Jet Ridge Line M.I. Example]{Radio maps at 5 GHz of 1928+738 (left) and 0700+470 (right), with jet ridge lines superimposed. Two examples of a sinusoid-like jet morphology (with M.I.=0.75, left) and a single-bent one (with M.I.=0, right). VLBI maps from \citet{Britzen2007a}.}
  \label{fig:ridge_lines_example}
\end{center}
\end{figure*}

\subsection{Monotonicity Index, M.I.}
\label{sec:rl.MI}

Detailed VLBI investigations in the past have shown that a large number of active galaxies exhibit bent or otherwise non-linear jet morphologies on various scales. Individual sources like S5 1803+784 and PKS 0735+178, as well as others (e.g. 3C 345; \citealt{Lobanov1999}; B0605-085; \citealt{Kudryavtseva2010}), have been extensively studied to understand the origin of this bending. In this context, we are interested in quantifying the bending of the whole jet ridge line. Moreover, we want to differentiate between a ``monotonically-bent'' jet, i.e., a jet that is bent in only one direction (see Fig. \ref{fig:ridge_lines_example}, right), as opposed to a more sinusoid-like morphology (similar to what is seen for some epochs of S5 1803+784, see Fig. \ref{fig:ridge_lines_example}, left). This is done through the use of the monotonicity index, M.I..

We quantify a sinusoid-like morphology of a jet by identifying the local extrema in a given jet ridge line, in the core separation - position angle plane. For an epoch \emph{i}, a component \emph{m} exhibits a local extremum under the definition
\begin{equation}
\theta_{m;extr}:|\theta_{m}-\theta_{m\pm1}|\geq 10(d\theta_{m}+d\theta_{m\pm1}),
\label{MI}
\end{equation}
where $\theta_{m}$ and $d\theta_{m}$ denote the position angle of component \textit{m} and its uncertainty. $\theta_{m}$ is calculated as
\begin{equation}
\theta_{m}=\arctan\frac{X}{Y},
\end{equation}
where \textit{X} and \textit{Y} are cartesian coordinates on the plane of the sky\footnote{We assume a conservative value of $10\sigma$ in Eq. \ref{MI} (where $\sigma$ here is defined as the sum of the position angle errors for $\theta_{m}$ and $\theta_{m\pm1}$) as the lower limit for flagging an extremum. This reflects an original underestimation of the individual $\theta_{m}$ uncertainties in the component fitting of \citet{Britzen2007a}.}.
Having calculated the number of extrema for a given jet ridge line at an epoch \textit{i}, we define the M.I. as
\begin{equation}
M.I.=\frac{number\;\;of\;\;extrema}{N-1},
\end{equation}
where \textit{N} is the total number of components at that given epoch. This is a crude calculation, but it can give us a measure of how the bending of the jet behaves along the jet. For M.I. values close to one, the jet more closely resembles a sinusoid. An M.I. value close to zero reveals a monotonic, single-bend, jet morphology. We normalize for the number of components \textit{N} to account for longer, or shorter, jets and to enable comparison between different sources. An example of such differences in the jet bending is shown in Fig. \ref{fig:ridge_lines_example}. 

\subsection{Apparent jet width, dP}
\begin{figure}[hbt]
\begin{center}
  \includegraphics[width=0.4\textwidth,angle=0]{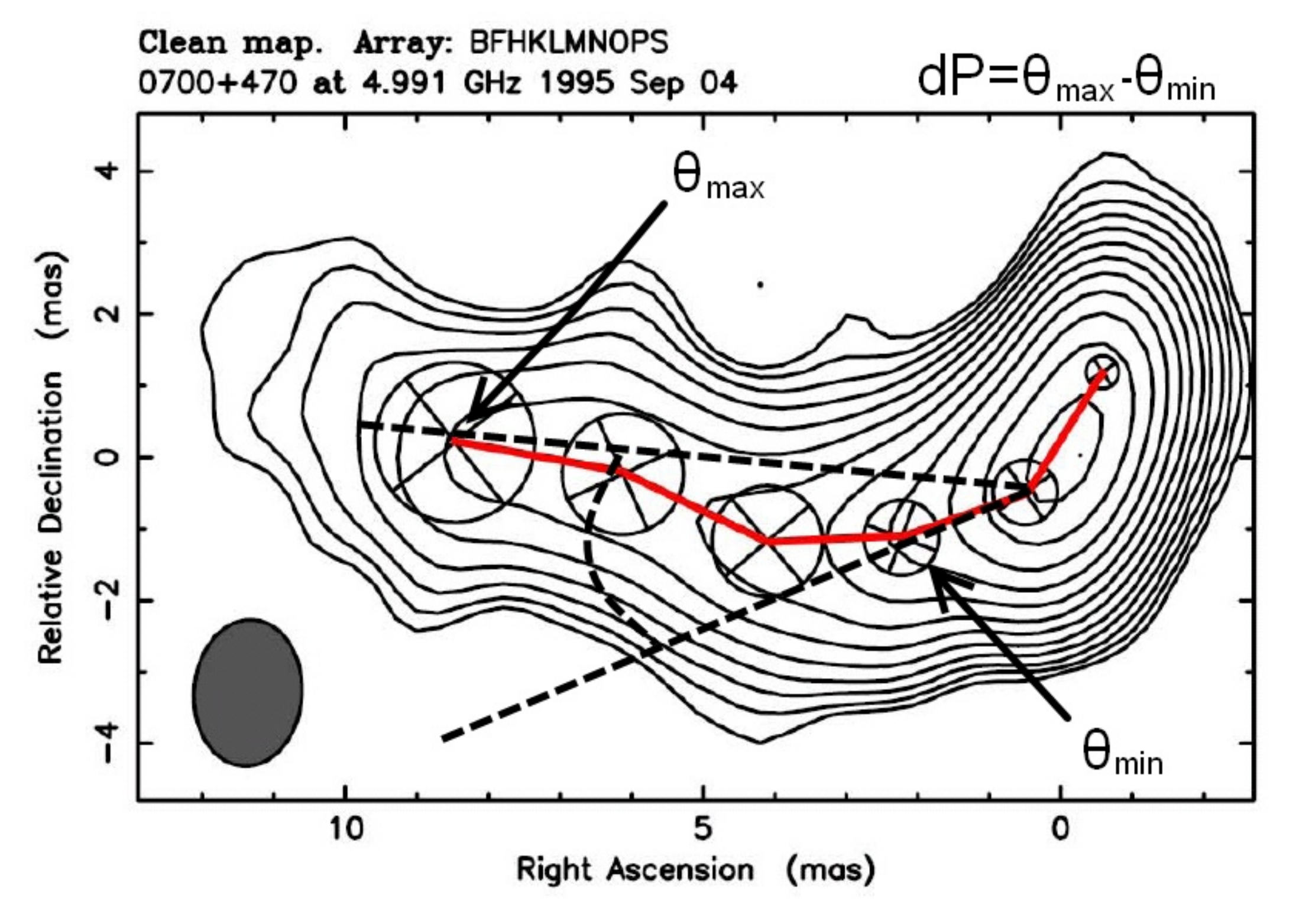}
  \caption{Radio map of 0700+470 at 5 GHz, exhibiting the calculation of the jet ridge line width. The two arrows denote the components at maximum and minimum position angle, while the area in between is defined as the opening, or apparent width, of the jet ridge line. The VLBI core is found at (0,0) coordinates. VLBI map from \citet{Britzen2007a}}
  \label{fig:ridge_line_width_example}
\end{center}
\end{figure}

For an epoch \emph{i} and a jet consisting of \textit{N} components characterized by their distance from the core and position angle $(r_{i},\theta_{i})$, we identify the components with the maximum and the minimum position angles. The apparent width of the jet dP, measured in degrees, is then calculated as (see Fig. \ref{fig:ridge_line_width_example}):
\begin{equation}
dP_{i}=\theta_{i}^{max}-\theta_{i}^{min},
\end{equation}
while the error is calculated by the propagation of errors formula.
One obvious drawback for the above definition is the non-localized nature of this measure. Using two different components, at different core separations, gives us only an approximate notion of the width of the flow. By using the definition above, we seek to quantify the opening of the jet flow and identify the effect of a channel-like jet, as seen in the case of S5 1803+784. 

\subsection{Apparent jet linear evolution, $\Delta\ell$}
The linear evolution across all available epochs and for all components of a jet can be calculated, ultimately producing a value that reflects the total linear displacement of the whole jet ridge line. We use plane-of-the-sky coordinates $(X_{i},Y_{i})$ to calculate the linear displacement of component \emph{m }between epochs \emph{i} and \emph{(i+1)}:
\begin{equation}
l^{m}_{i}=\sqrt{\Delta X^{2}+\Delta Y^{2}}.
\end{equation}
To calculate the total displacement of the whole jet ridge line we then need to sum up over all components and all available observing epochs:
\begin{equation}
\ell=\sum_{i-1}\sum_{m}l^{m}_{(i)}.
\end{equation}
We need to account for both the different time span of observations, as well as the different number of components. We therefore define the jet linear evolution $\Delta\ell$, measured in parsecs per unit time and per component, as follows:
\begin{equation}
\Delta\ell=\frac{\sum_{i-1}\sum_{m}l^{m}_{(i)}}{{N(T_{i}-T_{1})}}=\frac{\ell}{NdT},
\end{equation}
where \textit{N} here is the total number of components used across all the epochs.
For the calculation of the displacement of an individual component $\ell$ between two consecutive epochs, and consequently of $\Delta\ell$, the cross-identification of components across epochs is necessary. Therefore, for $\Delta\ell$, unlike the previously discussed jet ridge line characteristic values, the actual identification of components is important.

$\Delta\ell$ essentially reflects the apparent speed distribution of all cross-identified components of the jet and therefore represents a value characteristic for the whole jet, rather than for any individual component. By summing up all components and epochs we trade temporal and positional resolution for a universal treatment of the entire jet. In this way we can test whether the kinematics of BL Lac objects is fundamentally different than that of FSRQs while averaging out localized properties of individual components. $\Delta\ell$ can be seen as a mean jet component speed, with the difference that it is acquired through averaging not only over all components, but additionally over all available epochs. Although it certainly reflects a measure of the outward motion in BL Lac jets, the calculation of $\Delta\ell$ is done in such a way that the potential curvature of the components' trajectories is taken into account. This separates $\Delta\ell$ from a simple linear regression fit to the core-separation versus time diagrams usually employed to calculate outward velocities, making it sensitive to non-radial motions, that are otherwise missed.\\

\section{Analysis and results}
\label{sec:analysis}
Since the CJF sample spans a large redshift range, combined with the fact that we are using apparent quantities, i.e., measures that are projected on the sky, different linear scales will be probed at different redshift ranges. To avoid this effect we impose two constraints to our data.
\begin{itemize}
\item As most CJF sources have available spectroscopic redshifts, we can calculate their projected jet length in linear scales. We can then impose a constraint to the maximum core separation out to which a component is considered in this investigation. Thus we limit ourselves to the same linear scales for all of the CJF objects, albeit with different linear resolution. As we are interested in the behavior of BL Lacs, we use them as the basis of our decision. For S5 1803+784, peculiar kinematics of the components is observed for the inner-most part of its jet, out to 6 mas. For this source's redshift (z=0.68), this translates roughly to 41 pc. Similar to S5 1803+784, PKS 0735+178 and S5 0716+714 also show the most prominent evolution of their jet ridge line out to $\sim6$ mas (39.6 pc and 33 pc, for their respective redshifts). Given that most of the BL Lacs in our sample are at $z<1$, the length scale implied by the three sources mentioned above (the only objects whose jet ridge line has been investigated in detail) ensures that the inner-jet of sources at z $<1$ is covered. We adopt therefore an apparent core separation limit of 40 parsecs.
\item As we are interested in the behavior of BL Lac objects, we can limit our comparisons to the redshift range [0,1]. In this way we can partially account for the different linear scales, as well as avoiding any possible evolution of the values we are investigating as a function of redshift. There are 21 BL Lac sources with available redshift information, with 17 of them being at z $<1$. There are 56 FSRQs in the same redshift range.
\end{itemize}

\subsection{Jet ridge line morphology}
We investigate how, if at all, a jet resembles a sinusoid (as in the case of S5 1803+784). In order to do this we calculate the M.I. from the jet ridge lines of our sources. The M.I. is calculated for the whole jet (we do not apply the core separation limit of 40 pc), as we are interested in the morphology of the entire jet ridge line, rather than a localized property. Additionally, a large number of components is needed for a robust interpretation of the M.I., we therefore take into account only epochs with at least three components identified. Finally, we define the maximum M.I. value for each source across the available epochs. Fluctuations of the M.I. of a source between epochs are low.

\begin{figure}[h]
\begin{center}
  \includegraphics[width=0.45\textwidth,angle=0]{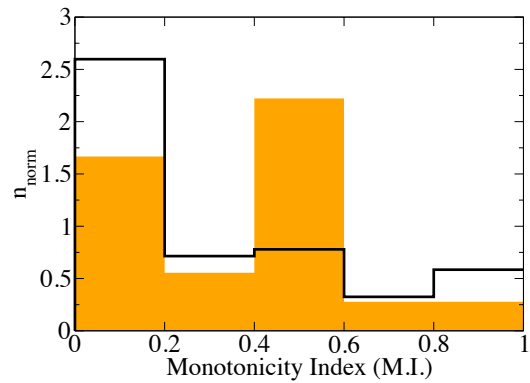}
  \caption{Monotonicity Index (M.I.) distribution for BL Lacs (filled line) and FSRQs (open line). The histogram has been normalized to area unity.}
  \label{fig:MI_types_histo}
\end{center}
\end{figure}

Under the above constraints, we find that there are in total 46 sources with M.I. $\geqslant0.5$. This translates to 37\% of the total number of sources used here. BL Lacs show more often jets with sinusoid-like morphologies, compared to FSRQs. More than half of the FSRQs ($\sim 51\%$) have an M.I. value of 0, while two thirds ($\sim 66\%$) have an M.I. value lower than 0.5. In contrast to this, only one third of the BL Lacs ($\sim 33\%$) have M.I.=0 and less than half ($\sim 44\%$) of them have M.I.$<0.5$. Figure \ref{fig:MI_types_histo} shows the histogram of the M.I. for the different types of objects. BL Lacs show their maximum in the [0.4,0.6) bin. In contrast, FSRQs have their maximum in the [0,0.2) bin. When considering the median values for the two classes of objects, BL Lacs and FSRQs differ at $>3\sigma$ level ($0.500\pm0.014$ compared to $0.000\pm0.004$, respectively). It should be noted that a robust quantitative treatment of the M.I. is problematic, given the severe restrictions imposed by the small number of components per jet, as well as the limited temporal resolution of the observations used here.

BL Lacs appear to more often show sinusoid-like curved jets, as seen for S5 1803+784 and 0716+714.

\subsection{Apparent jet width, \textit{P}, statistics}
We calculate non-zero jet widths for a total of 559 epochs (multiple epochs per source). For one epoch of the source S5 1803+784 and one epoch of the source 2200+420, a width of $\sim180$ degrees is found. This is probably due to a problem in the original component fitting and we therefore exclude these two epochs in this analysis. Of the remaining 557, 277 are larger than 10 degrees, while 125 are larger than 20 degrees. This translates to 22.4\% of the sample possessing jets with widths exceeding 20 degrees.

We calculate an average jet ridge line width for our\linebreak sources. We find that BL Lacs show significantly wider jet ridge lines (average value of $19.3^{\circ}\pm1.8^\circ$), compared to FSRQs ($12.0^{\circ}\pm0.4\circ$; $4\sigma$ difference). A Student's t-test gives a significance $>$99.99\% that BL Lacs and FSRQs show different mean values. As discussed above, we also investigate sources in the redshift bin [0,1]. In Table \ref{tab:jet_width_statistics_z_1} we show the characteristic statistical parameters for the CJF jet ridge line widths for all sources with available redshift, as well as sources in the redshift bin [0,1]. 

\begin{table}[htb]
\begin{center}
\caption{Characteristic statistical values concerning the width of the jet ridge line of sources with measured redshifts. We give average and median values with uncertainties for FSRQs and BL Lacs.}
\label{tab:jet_width_statistics_z_1}
\begin{tabular}{l | l l | l l}
\multicolumn{5}{l}{Jet Width ($^\circ$)}\\
\hline
 &\multicolumn{2}{c}{All} &\multicolumn{2}{c}{ $\mathbf{0 < z < 1}$}\\
\hline						
\textbf{Types}	&\textbf{FSRQ} &\textbf{BL} &\textbf{FSRQ}	&\textbf{BL}\\
\hline	\multicolumn{5}{l}{} \\
\textbf{\#}	                &343		&63	&91	&58\\
\textbf{Average} &12.0	&19.3	&11.1	&20.4\\
\textbf{Error}	 &0.4	      &1.8	&0.8		&2.0\\
\textbf{Median}	 &8.7   &12.5	&7.5		&13.8\\
\textbf{Error}      &0.5       &2.2	&1.0		&2.5\\
\multicolumn{5}{l}{}\\
\hline
\end{tabular}
\end{center}
\end{table}

Focusing on the redshift bin [0,1], from Table \ref{tab:jet_width_statistics_z_1} we see that BL Lac jet ridge lines appear significantly wider than their FSRQ counterparts ($20.4^{\circ}\pm2.0^\circ$, compared to $11.1^{\circ}\pm0.8^\circ$; $>4\sigma$). The same behavior is seen when looking at the median values (albeit at the $2\sigma$ level).

\begin{figure}[htb]
\begin{center}
   \includegraphics[width=0.445\textwidth,angle=0]{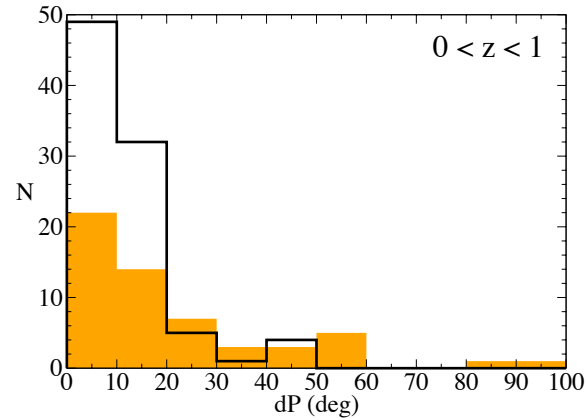}
  \caption{Jet ridge line width distribution for BL Lacs (filled line) and FSRQs (open line). For this histogram sources with redshifts in the [0,1] bin are used.}
  \label{fig:jet_width_types_0_z_1}
\end{center}
\end{figure}

In Fig. \ref{fig:jet_width_types_0_z_1} we show the distribution of jet ridge line widths for BL Lacs and FSRQs, in the redshift bin [0,1]. Both\linebreak classes show similar distributions, with their maxima situated at around 10 degrees. The distribution of BL Lac appears to be wider and extending to larger widths, compared to FSRQs. FSRQs show a rather more contained distribution to lower width values than BL Lacs. For the 125 epochs where CJF sources show apparent jet widths $>20$ degrees, we find 14 BL Lacs and 27 FSRQs. When taking into account the total number of each type of object, we calculate that 47.3\% of BL Lacs have jets wider than 20 degrees, as opposed to only 13.6\% for FSRQs.

We apply the two-sample Kolmogorov-Smirnoff (K-S) test to our data to see whether the BL Lacs and FSRQs are indeed different in their jet ridge line widths. Comparing the jet ridge line widths of BL Lacs and FSRQs (no redshift binning) the K-S test gives a $0.7\times10^{-3}$\% probability that these two sub-samples originate from the same parent distribution. We also compare the sub-samples of BL Lacs and FSRQs in the redshift bin [0,1]. In this case the K-S test gives a probability of 1\% that BL Lacs and FSRQs are drawn from the same parent sample. We conclude that BL Lacs show a significantly different distribution of jet ridge line widths compared to FSRQs. In both average and median values, BL Lacs show substantially wider jets than FSRQs.

\subsection{Apparent jet linear evolution, $\Delta\ell$, statistics}
Focusing on the final measure of the jet ridge line kinematics, as studied in this paper, we want to investigate whether the stationarity of components, as observed in the case of S5 1803+784, is commonplace among other BL Lacs. We use the linear evolution measure, as described in Sect. \ref{sec:data}, to do this. We follow the same procedure as previously to check the statistics of the individual classes.

In Table \ref{tab:jet_width_ev_max_z} we give the statistical properties of the total linear evolution of the jet ridge line (measured in parsecs per unit time and component) distributions for FSRQs and BL Lacs. On average, BL Lacs show weaker evolution of their jet ridge lines compared to FSRQs ($\sim2\sigma$ difference). Looking at the median values, BL Lacs show the least evolution compared to FSRQs ($0.24\pm0.06$ and $0.414\pm0.020$ pc/yr/comp respectively; $\sim3\sigma$ difference). We also study the statistics of the sub-sample of CJF sources in the [0,1] redshift bin (also in Table \ref{tab:jet_width_ev_max_z}). The behavior remains the same as before, with differences between BL Lacs and FSRQs becoming more pronounced (with a $\sim4\sigma$ difference in median values).

\begin{table}[htb]
\caption{Statistical properties of the jet linear evolution distributions for FSRQs and BL Lacs. Average and median values are calculated for both all sources, as well as for sources in the redshift bin [0,1]. Only the inner part of the jet is considered ($<40$ pc).}
\label{tab:jet_width_ev_max_z}
\begin{center}
\begin{tabular}{l | l l | l l }
\multicolumn{5}{l}{}\\
\multicolumn{5}{l}{\large{\textbf{Jet linear evolution} ($pc/yr/comp$)}}\\
\hline
	&\multicolumn{2}{c|}{$0 < z$}	&\multicolumn{2}{c}{$0 < z < 1$}\\			
Types &\textbf{FSRQ}	&\textbf{BL}	&\textbf{FSRQ}	&\textbf{BL}\\
\hline\multicolumn{5}{l}{}\\
\textbf{\#} &171  			&25   					&44           		    &21         	      \\
\textbf{Average} &0.470		&0.38					&0.51 		    &0.38 		      \\
\textbf{Error} &0.016		&0.05					&0.04 		    &0.05		      \\
\textbf{Median} &0.414		&0.24					&0.50       	  	    &0.24     	      \\
\textbf{Error }&0.020      		&0.06       		           		&0.04           	    &0.06       	      \\
\multicolumn{5}{l}{}\\
\hline
\end{tabular}
\end{center}
\end{table}

\begin{figure*}[htp]
\begin{center}
  \includegraphics[width=0.5\textwidth]{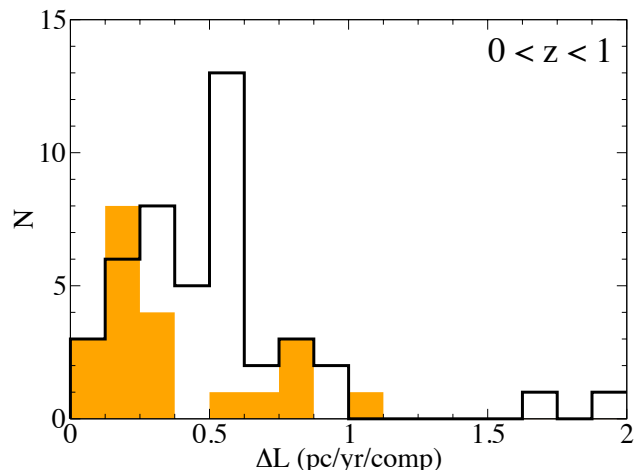}
  \caption{Distribution of the total linear evolution of the CJF jet ridge lines for FSRQs (open line) and BL Lacs (filled line). We use sources with redshift lower than 1.}
  \label{fig:jet_linear_ev_histo}
\end{center}
\end{figure*}

We now turn to the actual distribution of the total jet linear evolution for the three types of objects, shown in Fig. \ref{fig:jet_linear_ev_histo}. We once again only take into account sources in the redshift bin [0,1]. Quasars show a pronounced maximum around 0.63 pc/yr/comp, extending out to 2 pc/yr/comp. In contrast, BL Lacs show their distribution maxima around 0.25 pc/yr/comp. Quasars appear to have a wider distribution of total linear evolution values, also showing the highest maximum values among all three types (see Table \ref{tab:jet_width_ev_max_z}). Once again we employ the K-S test to compare the distributions. For the whole sample (independent of redshift constraints), the K-S test gives a probability of 4.3\% that BL Lacs and FSRQs are drawn from the same parent population. Focusing on the $0<z<1$ sub-sample, we get a marginal 5.4\% probability that BL Lacs and FSRQs stem from the same parent sample. These results give a positive answer as to whether BL Lacs show less absolute linear evolution of their jets compared to their FSRQ counterparts.

\section{Discussion}
\label{sec:discussion}

\begin{figure}
\begin{center}
\includegraphics[width=0.4\textwidth]{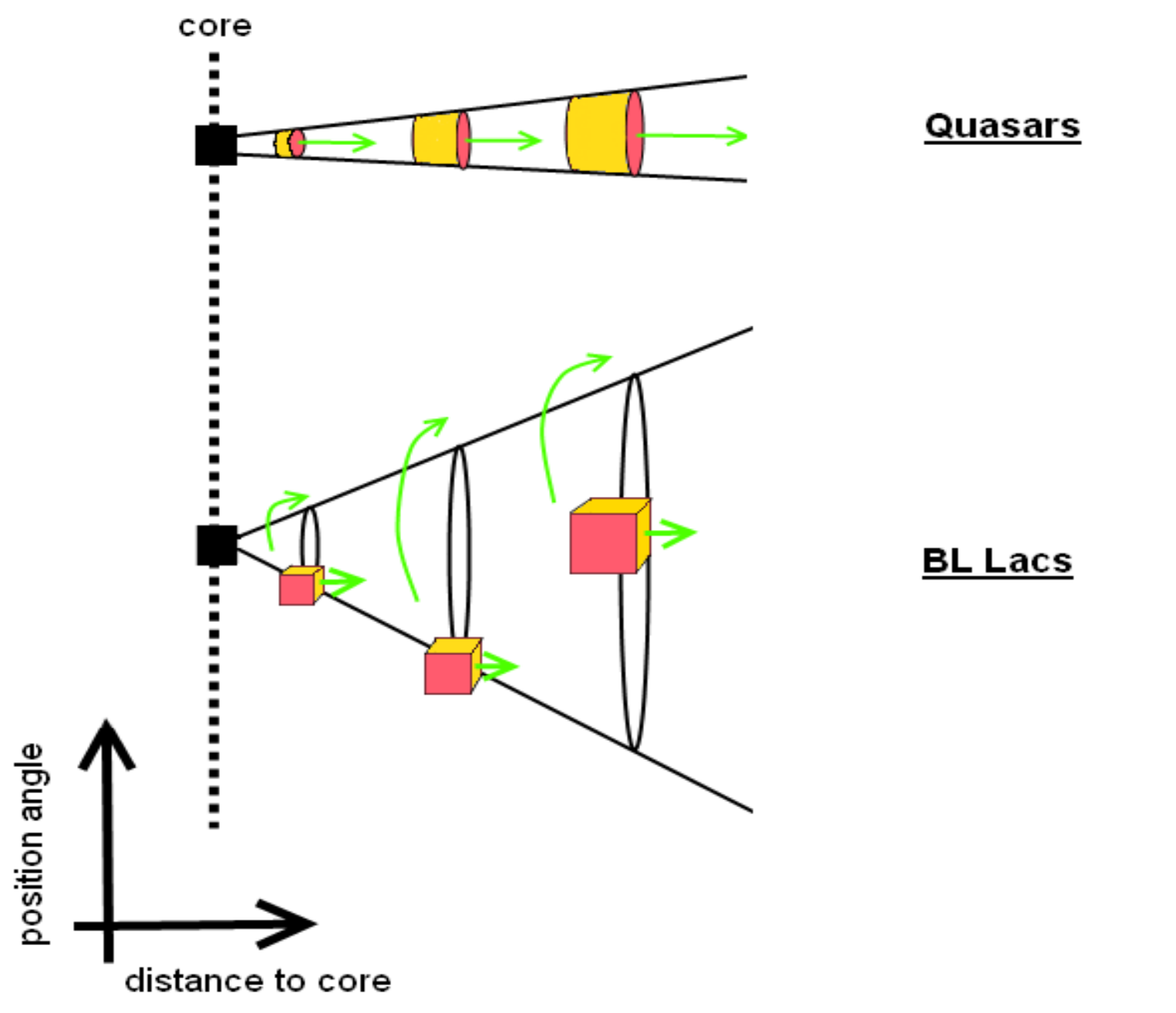}
\caption{A sketch of the new paradigm for jet kinematics in BL Lac objects, compared to the established scheme of superluminal motion in blazars. BL Lac jet components appear to move less outwards, while exhibiting stronger non-radial motion components.}
\label{fig:newparadigm}
\end{center}
\end{figure}

Figure \ref{fig:newparadigm} shows a sketch of the new paradigm as was originally seen in the BL Lac object S5 1803+784 and the effects of which we tried to investigate in the previous sections. Our statistical investigation, part of which is presented here, lends support to a different kinematic scheme for BL Lac objects. Indeed we find that BL Lac jet components show significant non-radial components in their motion,\linebreak compared to those of FSRQs. It should however be noted that there is a substantial part of the total CJF sources that show similar, ``BL Lac-like'' behavior. This implies the presence of a more universal effect than simply a property characteristic to BL Lacs.

\subsection{Possible explanations}
It is of great interest to try and explain the kinematic behavior seen in a large number of the CJF sources. Combining the percentages calculated from both apparent width and apparent linear evolution we get an approximately 30\% of the sample showing wide jets that show slow outwards motions. There also seems to exist an apparent preference for BL Lacs to show this behavior over FSRQs. One obvious factor that should play a deciding role for the kinematics of jet components is the viewing angle under which a source is observed. BL Lacs and FSRQs are assumed to be seen jet-on, at the smallest viewing angles, with steeper-spectrum quasars and radio galaxies being sources observed at progressively larger viewing angles (e.g., \citealt{Antonucci1993};\linebreak \citealt{Urry1995}). Could a viewing angle difference explain the kinematic differences seen in some of the CJF sources and in particular observed between BL Lacs and FSRQs? Assuming that jet components follow ballistic, linear paths, one can expect that viewed at smaller viewing angles (smaller than the critical angle $1/\gamma$, with $\gamma$ being the Lorentz factor of the flow), the components are observed to cover smaller distances than if seen edge-on. Therefore, the slower speeds, and hence smaller total linear evolution of their jet ridge lines could be explained in terms of a systematically smaller viewing angle for BL Lacs compared to FSRQs.

However, \citet{Hovatta2009} used long timescale variability of AGN, together with jet kinematics, to derive\linebreak Doppler factors and consequently viewing angles of a sample of 87 AGN. They found that FSRQs actually show\linebreak smaller mean viewing angles than BL Lacs (a result also found by \citealt{Laehteenmaeki1999}). Although there are certain limitations to the method used by the authors, their results are difficult to reconcile with a viewing-angle-dependent explanation of our results. 

Our results would therefore indicate the need for an additional, potentially geometric, effect in play. This would be in agreement with that low-speed components in FSRQ and BL Lac jets should appear so because their pattern Lorentz factor is lower than the bulk Lorentz factor of the jet (e.g., \citealt{Cohen2007}). Alternatively, it should be considered whether there is some systematic bias in the way Doppler factors are estimated that would lead to an over- or underestimation of the viewing angles of BL Lacs and FSRQs, respectively.

The large jet widths are more difficult to explain. An additional mechanism or effect must be introduced to produce a wide jet. Precession of the jet axis, or assuming that the components follow non-ballistic and non-linear paths, can lead to such jet properties (e.g., \citealt{Steffen1995}; \citealt{Gong2008}; \citealt{Roland2008}; \citealt{Gong2011}). In these models, jet components follow highly curved trajectories, which, viewed face-on, would give the impression of a\linebreak wider distribution of jet component position angles. A precessing jet, or rather a precessing jet nozzle, would imply that different components follow different trajectories. That would result in a changing jet component position angle distribution, more pronounced at smaller viewing angles. 

That more BL Lacs show M.I. closer to one compared to FSRQs, implying a highly curved, sinusoid-like ridge-line, lends additional support to this scenario. That leads us to the conclusion that either (1) there is a sub-set of FSRQs and BL Lacs ($\sim30\%$ in our sample) for which non-ballistic and/or precession effects play an important role, or more generally that (2) for these sources the viewing angles are below some critical angle $
\theta_{c}$ that allow us to witness and thus characterize the helical or ``non-ballistic'' structure common in all radio-AGN jets, but which would otherwise be blended out by the projection effects at viewing angles $>\theta_{c}$.

A final scenario that needs to be discussed is whether a systematic difference in the Lorentz factors $\gamma$ between \mbox{FSRQs} and BL Lacs could produce the effects observed here. It has been argued that BL Lacs show smaller Lorentz factors than FSRQs (e.g., \citealt{Morganti1995};\linebreak \citealt{Urry1995}; \citealt{Hovatta2009}). If this is true (and not a selection bias in the samples used) then that\linebreak would naturally explain the slower components in BL Lacs. 
The above in turn ties in to the currently accepted unification scheme (e.g., \citealt{Urry1995}). In that paradigm, BL Lacs and \mbox{FSRQs} are drawn from two, presumably different, parent samples of Fanaroff-Riley I (low luminosity) and II (high luminosity) galaxies (FR; \citealt{Fanaroff1974}; \citealt{Padovani1991}; \citealt{Capetti1999}; \citealt{Xu2009}), respectively. If that is true, the differences seen in the jet ridge line properties of BL Lacs and FSRQs should then translate to differences between FRI and FRII jet kinematics. Interestingly, studies of FRI and FRII kinematics have shown that both types have similar parcec-scale jet Lorentz factors (e.g., \citealt{Giovannini2001}), contradictory to what we find here, as well as in other studies mentioned above, concerning the parsec-scale jet speeds in BL Lacs and FSRQs. 

\subsubsection{Helical jets}
There is a growing number of sources for which helical models are employed to explain their kinematic, flux, and evolution properties (e.g., 3C 273, \citealt{Lobanov2001}; 1803+784, \citealt{Roland2008}; 3C454.3, \citealt{Qian2007}; 0605-085, \citealt{Kudryavtseva2010}; 1632+382, \citealt{Liu2010}, etc.). A helical, bent, trajectory has been employed for 3C345, showing that a constant Lorentz factor of about 10 can explain the observations. Apparent acceleration of the components is an effect of the change of the viewing angle (\citealt{Zensus1995}). Further evidence supporting an intrinsic helical structure for AGN jets (as traced by the magnetic fields threading them) is given by polarization studies, and in particular Faraday rotation gradients and circular polarization found in several sources (e.g., \citealt{Gabuzda2005}; \citealt{Gabuzda2008}).

We can use a simple helical jet model and try to see whether a helical geometry, combined with projection effects, can account for the results of the statistical analysis above, namely that BL Lacs are expected to have wider jets,  than FQSOs, and also show weaker total linear evolution of their jets.

\begin{figure}[htb]
\begin{center}
  \includegraphics[width=0.3\textwidth,angle=0]{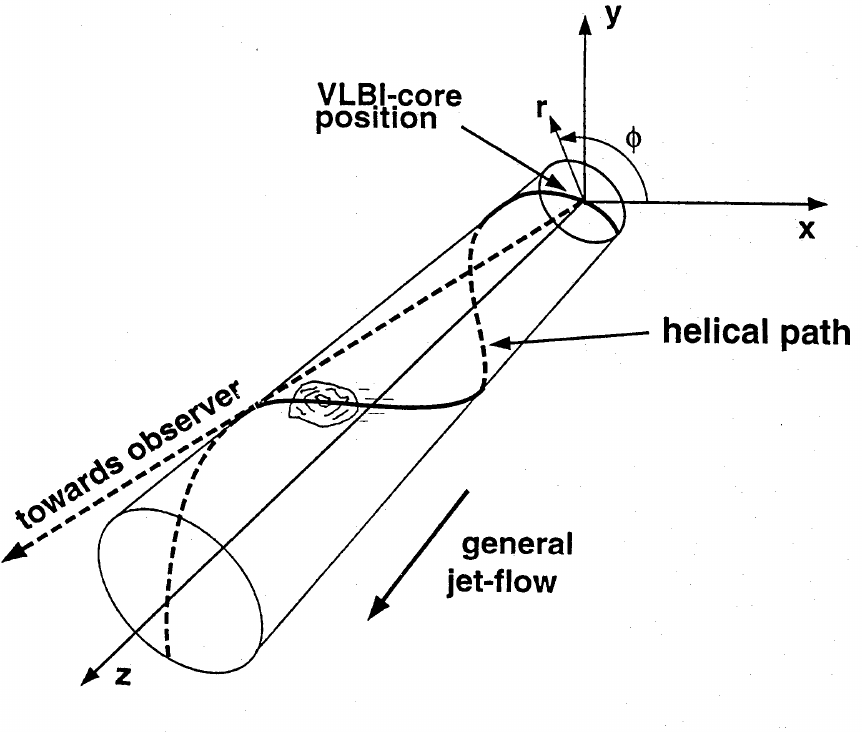}
  \caption[Geometry of the helical jet model]{The geometry of the helical jet model of \citet{Steffen1995}. The VLBI jet core is put at the origin of the cylindrical coordinates system (r, $\phi$, z). The jet axis (parallel to the z-axis) forms an angle $\theta$ to the line of sight. Figure reproduced from \citet{Steffen1995}.}
  \label{fig:steffen_geometry}
\end{center}
\end{figure}

The helical model of \citet{Steffen1995}, applied in that case to the QSO 3C345 (the same model has also been applied to the BL Lac source S5 1803+784, \citealt{Steffen1995b}; QSO B3 1633+382, \citealt{Liu2010}) can be employed to that end. The motion of a component down the jet is determined by the conservation of a number of quantities (e.g., specific kinetic energy, specific angular momentum, etc.). In particular, \citet{Steffen1995} find that the best fit to 3C345 provides a case where the specific kinetic energy of the component, the specific angular momentum, as well as the opening half-angle of the jet are conserved (Case 3 in that paper). The geometry of the model can be seen in Fig. \ref{fig:steffen_geometry}.

\begin{figure*}[htb]
\begin{center}
  \includegraphics[width=0.7\textwidth,angle=0]{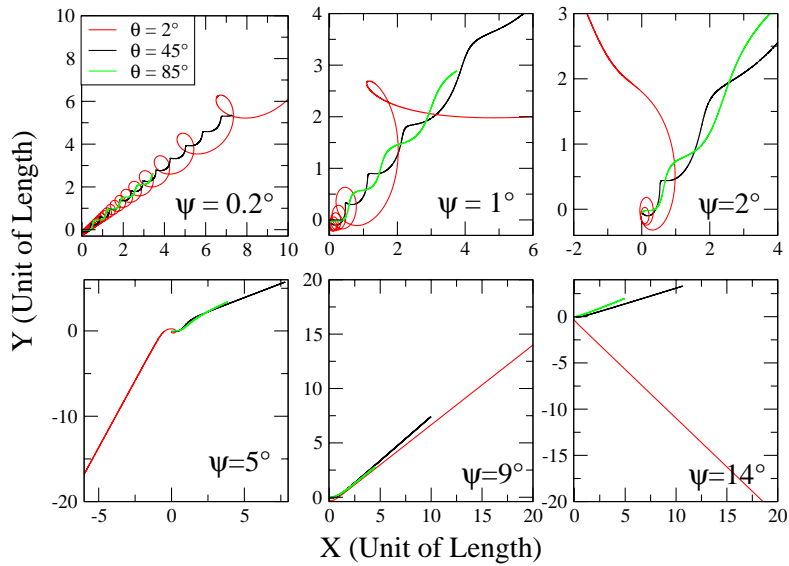}
  \caption[Helical trajectories examples]{Examples of different component trajectories for different opening half-angles: (a) $\psi=0.2$, (b) 1, (c) 2, (d) 5, (e) 9, and (f) 14 degrees, as well as three different inclination angles $\theta$ of the jet axis: (red) 2, (black) 45, and (green) 85 degrees. X and Y are sky coordinates, the plane perpendicular to the observer's line of sight.}
  \label{fig:steffen_examples}
\end{center}
\end{figure*}

In this context, the trajectory of a single component in the jet can be defined by the following set of parameters: the Lorentz factor and angular momentum of the component, the opening half-angle of the jet $\psi$, and the inclination angle of the jet axis $\theta$. Of particular interest for this investigation is the opening half-angle of the jet $\psi$, which defines how helical or linear the trajectory of a component is and the inclination angle of the jet axis, or viewing angle, $\theta$, which defines the projection effects at play. In Fig. \ref{fig:steffen_examples}, examples of component trajectories (in sky coordinates) are given for 6 different values of $\psi$. For each $\psi$ we allow for three different values of the viewing angle $\theta$. Both helicity and projection effects can be seen. A relativistic component with a speed $u=0.995c$ is assumed.

We can use this model to study whether and how the projected width of a jet depends on its helicity and the angle at which it is viewed. By definition, this model calculates the evolution of a single component. This is of course not the case of what is observed for most sources. The width of the CJF jet ridge lines is calculated as the difference (in degrees) between the components with the maximum and minimum position angle at a given epoch. Making the simplifying - and potentially simplistic - assumption that all components follow the same path (therefore implicitly assuming that $\psi$, $\theta$, and \textit{u} do not change between different components), this is equivalent to finding the maximum and minimum position angle for one component across different epochs.

\begin{figure*}[bt]
\begin{center}
  \includegraphics[width=0.45\textwidth,angle=0]{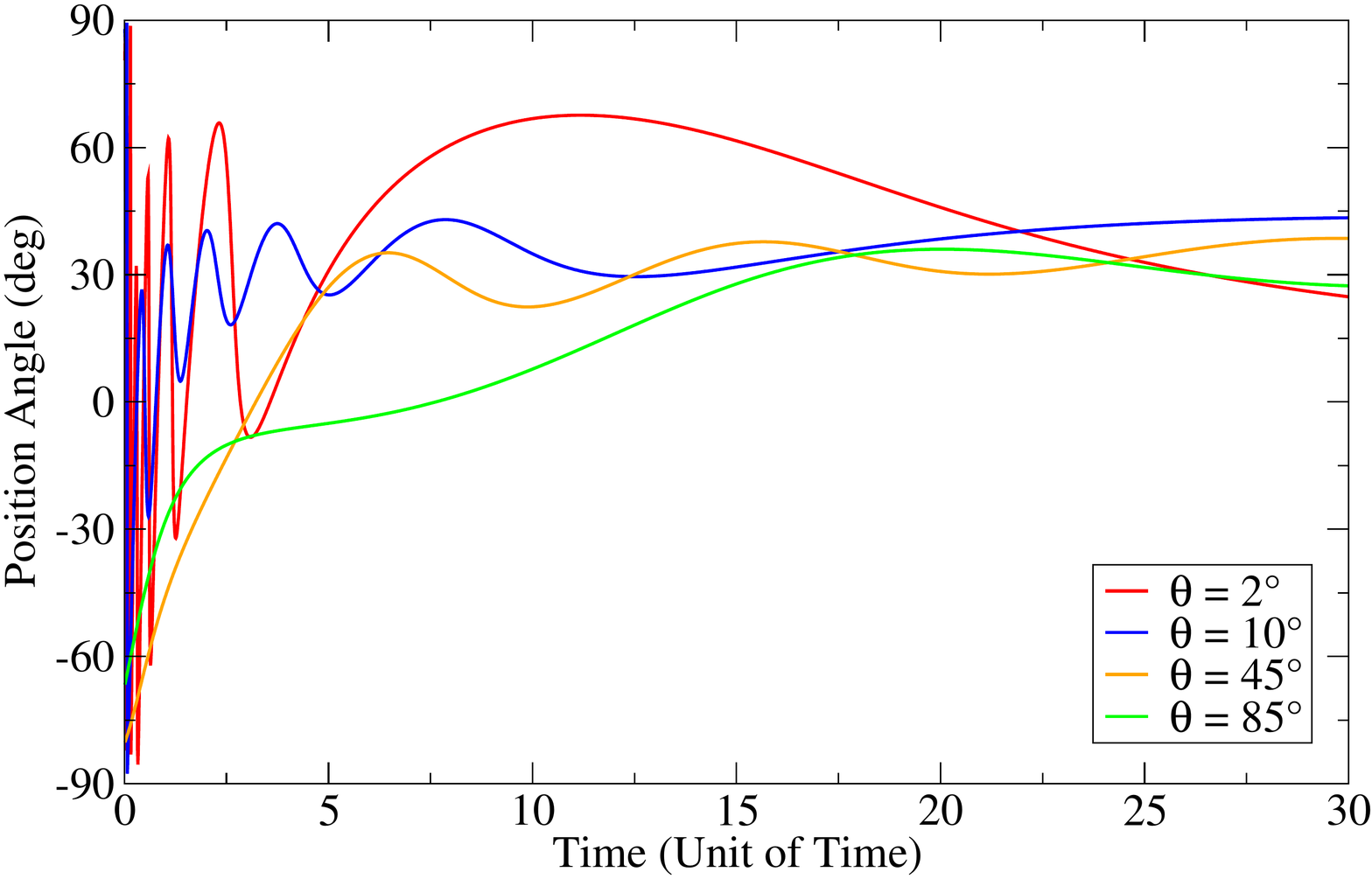}
   \includegraphics[width=0.45\textwidth,angle=0]{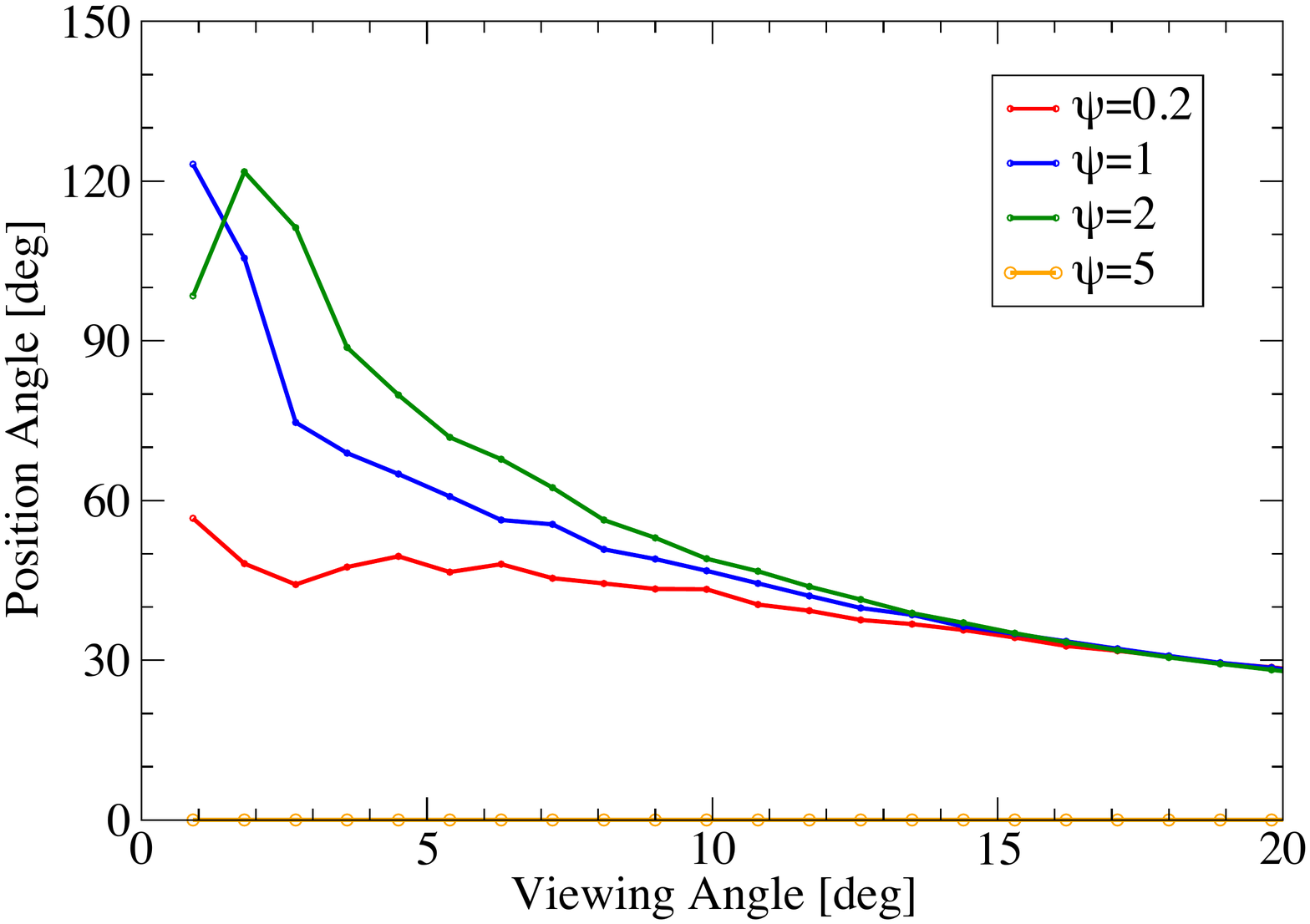}
 \caption[Position Angle Distribution with Time]{Left: Position angle as a function of time for a helical jet ($\psi=1$) and for 4 different viewing angles: (red) 2, (blue) 10, (orange) 45, and (green) 85 degrees. Right: Apparent jet width as a function of the viewing angle, for different opening half-angles: (red) 0.2, (blue) 1, (green) 2, and (orange) 5.}
  \label{fig:jet_width_steffen_example}
\end{center}
\end{figure*}

In Fig. \ref{fig:jet_width_steffen_example} (left) the evolution of the position angle of a component is shown as a function of time, for a helical jet ($\psi=1$) and for 4 different viewing angles. It is obvious that for smaller viewing angles, the modulations of the position angle become stronger, resulting in substantially wider position angle distributions than for the case of large viewing angles.

We can calculate the projected width of each loop of the helix. The maximum value of these widths (for a given viewing angle) is representative of that helix and therefore it is interesting to investigate its behavior as a function of $\theta$ and $\psi$. This is shown in Fig. \ref{fig:jet_width_steffen_example} (right). As expected, for highly helical jets (small values of $\psi$), smaller viewing angles produce substantially wider jets. The highest value of apparent width at the smallest viewing angle is produced by the $\psi=1$ jet, while for slightly larger viewing angles (up to $\sim15$ degrees) the $\psi=2$ jet appears wider. For $\psi>2$ zero apparent jet width values are not calculated. This is because of the method used to calculate the apparent width and reflects the fact that for these jets no single loop is produced in the rest-frame of the jet (i.e., these jets do not have a helical structure).

In conclusion, helical jets appear to reproduce well the effects seen for the width of the CJF jet ridge lines. A combination of a highly bent trajectory of the components, together with the effects of the projected geometry, can explain the extremely wide jets seen for BL Lacs (as it can be seen in Fig. \ref{fig:jet_width_steffen_example}, right). In this context, the subset of sources for which significant apparent jet widths are seen (i.e., dP $>20$ deg) are the sources at the smallest viewing angles. The trend for BL Lacs to show apparently wider jets than FSRQs would then imply that BL Lacs should be at systematically smaller viewing angles than the FSRQ sample. 

The origin of such a helically structured jet is an all together different question. Such helical structures/trajectories have been explained by a number of different models. An often employed concept is that of a precessing jet, combined with the non-ballistic motion of ejected components. For a detailed discussion see \citet{Britzen2010}.

%\begin{figure}[htb]
%\begin{center}
  %\caption[Apparent Jet Width with Viewing Angle]{}
  %\label{fig:jet_width_steffen_vartheta}
%\end{center}
%\end{figure}

\subsection{Comparison with previous studies}
It should be noted that this is the first time the jet ridge lines of a sample of AGN this size have been explicitly and exclusively studied in a statistical manner. While the ridge line of a jet is not a new concept, so far only the jet ridge lines of individual sources (both galactic and stellar) have been studied (e.g., \citealt{Condon1984}; \citealt{Steffen1995}; \citealt{Lister2003};  \citealt{Lobanov2006}; \citealt{Britzen2009}; \citealt{Perucho2009}; \citealt{Lister2009c}; \citealt{Britzen2010}). The definition of the jet ridge line can differ from study to study. We define the ridge line of a jet at a certain epoch as the line that linearly connects the projected positions of all components at that epoch. Alternatively, one can use an algorithm to more directly extract the jet ridge line from the VLBI map of each epoch (e.g., \citealt{Pushkarev2009}).

The one obvious direct comparison that can be drawn is to the MOJAVE / 2cm sample (\citealt{Lister2005}; Paper I). As the work presented here is prototype in its conception, it is difficult to draw direct comparisons to the MOJAVE sample. The CJF sources appear to be significantly slower than what is seen in the MOJAVE sample (e.g., compare Fig. 7 from \citealt{Britzen2008} and Fig. 7 from \citealt{Lister2009c}). In the distribution of apparent speeds, for the CJF there is a maximum at around 4c and then a turn down, with very few sources above 15c. In contrast, the distribution for the MOJAVE sources shows a maximum at 10c with a fair number of sources showing speeds greater than 20c. This can be explained in terms of the higher temporal resolution of the MOJAVE sample (owning to the larger number of epochs per source) that might allow the detection of such faster motions. It is interesting to note that no clear distinction has been made in the MOJAVE sample between BL Lacs and FSRQs, probably due to the relatively small number of BL Lacs. \citet{Lister2009c} however do mention that BL Lacs are more often found to exhibit ``low-pattern speeds", i.e., components with slow motions, significantly smaller than others in the same jet.  Similar evidence has been previously found in different VLBI samples (e.g., \citealt{Jorstad2001}; \citealt{Kellermann2004}). Combined with the argument that BL Lac jets exhibit lower Lorentz factors, as a result of a possible correlation between intrinsic AGN luminosity and jet Lorentz factor (e.g., \citealt{Morganti1995}), the above are in agreement with CJF BL Lacs showing on average weaker linear evolution than FSRQs. It should be noted that the agreement with previous studies concerning the outward motions of BL Lacs objects supports the robustness and reliability of our method.

A final note concerning the comparison between our results and these of the MOJAVE sample pertains to the observations frequency of each sample. Unlike the CJF, the MOJAVE uses VLBI observations at 15 GHz, meaning that a potentially different regime of the jet is probed, compared to what is studied in the CJF. Two-zone models, such as the ``two-fluid'' model (e.g., \citealt{Sol1989}, \citealt{Pelletier1989}), predict a fast jet spine, embedded in a slower moving sheath. This would imply that the MOJAVE potentially probes deeper, thus faster, jet layers compared to what is seen in the CJF sample. Moreover, higher observation frequency also translates to higher angular resolution and\linebreak therefore to probing effects closer to the core and also enabling the detection of faster moving components. In this case, it would be interesting to apply the methods used in this paper for the MOJAVE dataset, as it is expected that the curvature of the jet (and therefore potentially the angular properties and evolution of the jet ridge line) should play an increasingly important role at closer core separations.

\section{Conclusions}

We develop a number of tools to investigate both the morphology and kinematics of the CJF jet ridge lines:
        \begin{enumerate}
            \item Monotonicity Index
            \item Apparent jet width
            \item Apparent jet linear evolution
        \end{enumerate}
                
\noindent We summarize our results:
\begin{itemize}
    \item Using the M.I., we find that BL Lacs jet ridge lines more often resemble a sinusoidal curve compared to FSRQs. In contrast, 2/3 of the FSRQs have an M.I. lower than 0.5 (indicating fairly monotonic jets).
    \item 22.4\% of the CJF sources have apparent jet widths larger than 20 degrees. 47.3\% of BL Lacs, over 13.6\% for FSRQs, have $dP>20$ degrees.
    \item BL Lacs exhibit substantially apparently wider jets than FSRQs. This effect persists for a smaller redshift range. This supports the effects seen in individual BL Lac objects (i.e., 1803+784, 0716+714, etc.).
    \item The distribution of apparent jet ridge line widths for BL Lacs appears to extend towards higher values, with \mbox{FSRQs} mainly contained at lower values. A K-S test indicates a $0.7\times10^{-3}\%$ probability that FSRQs and BL Lacs are drawn from a single parent population.
    \item BL Lac objects, on average, show weaker apparent linear evolution of their jet ridge lines compared to FSRQs. 
    \item We use a helical jet model to show that a helical geometry combined with a small viewing angle can explain the large widths seen in BL Lac objects.
\end{itemize}

In conclusion, by statistically analyzing the CJF sample, we provide detailed insight concerning the morphology and evolution of AGN jet ridge lines. The statistical investigation of the CJF sources lends independent support to the different kinematic scenario recently seen in a number of BL Lac objects (1803+784, \citealt{Britzen2010}; 0735+178, \citealt{Britzen2010b}). 25\%-30\% of the CJF sample show apparently considerably wide jets and strong apparent width evolution. BL Lac objects appear to deviate the strongest from the kinematic paradigm, widely accepted for blazars, of outward superluminally moving jet components. BL Lacs appear to evolve their jet ridge lines (with respect to the core) less than the other source classes, hence indicating a slower apparent flow in their jets. On the other hand, they show significantly apparent wider jets than FSRQs. We have argued that the combination of a helically structured jet (as a universal property of all AGN jets) and projection effects would produce the apparent kinematic properties observed here. Viewing a source at a small angle to its jet axis possibly allows us to uncover this peculiar kinematic behavior that would otherwise be inaccessible at larger viewing angles. This would then imply that BL Lac objects are preferentially seen at smaller viewing angles than FSRQs. Combined with the significant number of (non-BL Lac) CJF\linebreak sources showing evidence that support this different kinematic scheme imply a rather universal effect, rather than something unique to BL Lacs.

Finally, it should be noted that the above results underline the fact that the notion of linear, ballistic trajectories for AGN jet components usually employed until recently is a very crude approximation and, more often than not, deviates grossly from the reality. It is of great interest to uncover the physical process that leads to the properties of AGN jet ridge lines as studied in this paper.

%\begin{figure}
%\includegraphics[]{}
%\caption{}
%\label{}
%\end{figure}
 
%\begin{equation}
 % \label{useless-equation}
  %\end{equation}

\acknowledgements
M.K. acknowledges the support from the\linebreak Creative Research Initiative program, No. 2010-0000712, of the National Research Foundation of Korea (NRFK) funded by the Korea government(MEST). M.K. was supported for part of this research through a stipend from the International Max Planck Research School (IMPRS) for Astronomy and Astrophysics. M.K. was supported for part of this research through the Reimar-L\"ust stipend of the Max Planck Society. M.K. wants to thank Lars\linebreak Fuhrmann and Mar Mezcua for insightful discussions and comments that greatly improved this manuscript. This research has made use of the NASA/IPAC Extragalactic Database (NED) which is operated by the Jet Propulsion Laboratory, California Institute of Technology, under contract with the National Aeronautics and Space Administration. This research has made use of NASA's Astrophysics Data System Bibliographic Services. The Very Long Baseline Array is operated by the USA National Radio Astronomy Observatory, which is a facility of the USA National Science Foundation operated under cooperative agreement by Associated Universities, Inc.

\bibliographystyle{aa}
\bibliography{bibtex}

\end{document}